\documentclass[12pt]{article}
\usepackage{a4wide}
\usepackage{amssymb, amsmath,cite}
\usepackage[normalem]{ulem}
\usepackage[utf8]{inputenc}
\usepackage[colorlinks]{hyperref}
\usepackage[usenames,dvipsnames]{color}
\usepackage{xcolor}
\usepackage{graphicx}
\hypersetup{
     breaklinks=true,
    pdfstartview={FitH},    
    colorlinks=true,       
    linkcolor=blue,          
    citecolor=red,        
    filecolor=magenta,      
    urlcolor=blue,           
    anchorcolor=green,      
    linktocpage=true
}

\begin{document}
{\renewcommand{\thefootnote}{\fnsymbol{footnote}}
		
\begin{center}
{\LARGE Role of trans-Planckian modes in cosmology} 
\vspace{1.5em}

Arjun Berera$^{1}$\footnote{e-mail address: {\tt ab@ph.ed.ac.uk}}, Suddhasattwa Brahma$^{2}$\footnote{e-mail address: {\tt suddhasattwa.brahma@gmail.com}} and Jaime R. Calder\'on$^{1}$\footnote{e-mail address: {\tt jaime.calderon@ed.ac.uk}}
\\
\vspace{0.5em}
$^{1}$School of Physics and Astronomy, University of Edinburgh, \\
Edinburgh, EH9 3FD, United Kingdom\\ \vspace{0.3em}
$^{2}$Department of Physics, McGill University\\
 Montr\'eal, QC H3A 2T8, Canada\\
\vspace{1.5em}
\end{center}
}
	
\setcounter{footnote}{0}

\newcommand{\bea}{\begin{eqnarray}}
\newcommand{\eea}{\end{eqnarray}}
\renewcommand{\d}{{\mathrm{d}}}
\renewcommand{\[}{\left[}
\renewcommand{\]}{\right]}
\renewcommand{\(}{\left(}
\renewcommand{\)}{\right)}
\newcommand{\nn}{\nonumber}
\newcommand{\Mpl}{M_{\textrm{Pl}}}
\def\H{H_f}
\def\V{\mathrm{V}}
\def\e{\mathrm{e}}
\def\be{\begin{equation}}
\def\ee{\end{equation}}

\begin{abstract}
\noindent Motivated by the old trans-Planckian (TP) problem of inflationary 
cosmology, it has been conjectured that any consistent effective field theory 
should keep TP modes `hidden' behind the Hubble horizon, so as to prevent 
them from turning \textit{classical} and thereby affecting macroscopic 
observations. In this paper we present two arguments against the Hubble horizon being a scale of singular significance as has been put forward in
the \textit{TP Censorship Conjecture} (TCC). First, refinements of TCC are 
presented that allow for the TP modes to grow beyond the horizon while 
still keeping the de-Sitter conjecture valid. Second,  we show that TP 
modes can turn classical even well within the Hubble horizon, which, 
as such, negates this rationale behind keeping them from crossing it. 
The role of TP modes is known to be less of a problem in warm inflation,
because fluctuations start out usually as classical. This allows warm inflation to be more resilient to the TP problem compared to cold inflation. To understand how robust this is, we identity limits where quantum modes can affect the primordial power spectrum in one specific case.
\end{abstract}

\section{Introduction}
Inflation has been rightfully heralded as one of the most successful paradigms of describing the early-universe since it can solve several problems of standard big-bang cosmology -- namely the `horizon', `flatness' and `monopole' problems \cite{Guth:1980zm,Linde:1981mu,Starobinsky:1980te,Brout:1977ix, Albrecht:1982wi}. However, perhaps the crowning glory of inflation is its ability to explain the origin the macroscopic density perturbations one observes today, which lie at the origin of large scale structure, as vacuum fluctuations \cite{Starobinsky:1979ty,Mukhanov:1981xt}. In spite of this, there have been two lingering doubts cast onto the validity of inflation -- the trans-Planckian (TP) problem \cite{Martin:2000xs,Brandenberger:2000wr} and the initial singularity problem \cite{Borde:1996pt,Borde:2001nh}. The latter can easily be swept under the carpet as a problem for quantum gravity, which would serve as a quantum completion of inflation. However, the former problem arises when considering inflation as an effective field theory (EFT) on curved spacetimes, without requiring any specific inputs from quantum gravity.

The intuitive understanding of the TP problem goes as follows -- if we look at the physical wavelength of a classical perturbation mode at late times and blue-shift it backwards in time, due to the expansion of spacetime, one might end up with a physical wavelength that is smaller than the Planck length ($\ell_{\rm Pl}$). Turning the biggest success of inflation on its head, it might be argued that for sufficiently long periods of \textit{accelerated expansion}, one would have macroscopic perturbations originating from TP quantum fluctuations. This would imply that inflation needs to be valid up until energy ranges beyond the Planck scale as an EFT, which is clearly in conflict with our understanding of quantum gravity. At first sight, it might be argued that setting a Planck-scale cut-off is all that is required to get rid of this problem. However, for expanding backgrounds, imposing such cut-offs typically leads to a time-dependent Hilbert space, with the degrees of freedom increasing in time, resulting in complicated non-unitary evolution \cite{Weiss:1985vw, Jacobson:1999zk}. Therefore, one typically needs some other mechanism to make sure that such TP modes are not part of one's EFT on curved spacetimes.

One way proposed in the recent trans-Planckian censorship conjecture (TCC) is to ban all such TP modes from ever crossing the Hubble horizon \cite{Bedroya:2019snp}. This was done in the spirit of limiting the effect of TP modes from ever being part of our observations. Although there were no theoretical motivations for invoking the TCC, it was argued that the benefits of having such a principle would be immense. Even if there are TP modes in the system, any \textit{consistent} EFT, with an ultraviolet (UV) completion, would be such that it would prevent them from ever being part of our observations. In order to achieve this, the authors of \cite{Bedroya:2019snp} formulated the TCC as the requirement that a TP mode never crosses the Hubble horizon, and in the process, puts an upper bound on the amount of inflation \cite{Bedroya:2019tba}. Consequently, the same requirement can be used to put a bound on the lifetime of a meta-stable de Sitter (dS) phase of expansion \cite{Bedroya:2019snp}.

However, there is another way to argue against the TP problem. From this point of view, one would say that the linear perturbation theory one relies on in claiming that classical modes would be blueshifted to unacceptably high energies due to a long period of inflation, breaks down much before reaching this regime. In other words, if we look at a perturbation mode at late times and track its physical wavelength going backwards, one would find that linear perturbations grow to become larger than $1$ (in some suitable units) before the physical wavelength becomes smaller than or equal to $\ell_\text{Pl}$. This means that the universe is extremely inhomogenous and anisotropic on Planck scales and we cannot trust linear perturbation theory on those scales. In this way, one is forced to take refuge in a UV-complete theory taking over from inflation before reaching such length scales and hope that the TP problem gets resolved in this UV theory. At the very least, calculations like this have shown that the TP problem, as formulated for inflation specifically, does not even arise due to the breakdown of (linear) perturbation theory \cite{Kaloper:2018zgi, Tanaka:2000jw}. In fact, long back at the time when the TP problem was first introduced, it had been argued that in any interacting field theory, TP modes would not cause any case for worry. In other words, if one considers gravitational backreaction during inflation, then all the modes must either be in the Bunch-Davies (BD) vacuum, or in some state that would typically reduce to the BD one with a short few $e$-foldings, and would therefore have no significant effect on the dynamics of inflation \cite{Kaloper:2002cs}.

At this point, let us segue into defining what we precisely mean by a TP mode. If the universe is extremely inhomogeneous and anisotropic on Planck scales, then it would mean very little to have field decompositions into Fourier modes on those scales\footnote{Indeed, as claimed in \cite{Kaloper:2018zgi}, fluctuations cause much of these regions to turn into blackholes on such scales, effectively becoming a dynamical cutoff. Then, in some of these regions inflation cannot even start.}. On the other hand, a TP mode typically refers to energy scales when the physical momentum of a Fourier mode is of the $\mathcal{O}(\Mpl)$, \textit{i.e.} when quantum gravity effects are non-negligible. Thus, for the purposes of this paper, we would assume that a TP mode has a physical wavelength that is slightly larger than $\ell_\text{Pl}$ such that one can follow its evolution on cosmological spacetimes from thereon. This is merely a conceptual point to emphasize that we are classifying a mode to be a TP mode by going as far back as a mode-decomposition is valid in the sense of an EFT on curved spacetimes.

Coming back to the TCC, it is important to note that it had been invoked as a physical motivation for the dS conjecture \cite{Obied:2018sgi, Ooguri:2018wrx, Garg:2018reu, Andriot:2018mav}, the latter arising out of the stringy swampland \cite{Ooguri:2006in}. The swampland is the space of EFTs which cannot have consistent UV completions and any sensible EFT must satisfy the so-called `swampland conjectures' \cite{Palti:2019pca, Brennan:2017rbf} to avoid being in it (and, instead, be in the landscape). One of the most interesting swampland conjectures is the dS conjecture which states that metastable dS states cannot arise from a consistent UV-theory of gravity. Although motivated by the lack of a satisfactory construction of dS vacua in string theory, it has consequently been argued from other swampland conjectures \cite{Ooguri:2018wrx, Hebecker:2018vxz, Rudelius:2019cfh, Brahma:2019iyy} such as the swampland distance conjecture (SDC) \cite{Ooguri:2006in, Baume:2016psm, Klaewer:2016kiy, Landete:2018kqf, Grimm:2018cpv,Blumenhagen:2018nts}, which has been extensively tested in string theory \cite{Grimm:2018cpv, Blumenhagen:2018nts, Grimm:2018ohb,Lee:2018urn, Lee:2019tst, Joshi:2019nzi, Corvilain:2018lgw}, albeit the derivation is only valid in parametrically large regions of moduli space \cite{Banlaki:2018ayh, Junghans:2018gdb}. Accepting the TCC on face value, it can be shown that the (refined) dS conjecture follows from it without any restrictive assumptions. Therefore, as shown in \cite{Bedroya:2019snp}, assuming the TCC implies the existence of the dS conjecture. However, in this paper, we would like to study the opposite question -- assuming the validity of the dS conjecture, and more generally of the swampland, does it necessarily imply the TCC?

What we shall show is that this is not necessarily true. As shall be shown in Sec-2, it would become obvious that any (reasonable) upper bound on the phase of quasi-dS expansion would lead to a similar version of the dS conjecture. In other words, if one restricts an inflationary phase by an upper bound such as
\begin{eqnarray}\label{RefTCC}
	N < \,\mathcal{O}(1)\, \ln\(\frac{\Mpl}{\H}\)\,,
\end{eqnarray}
where $N$ is the number of $e$-foldings and $\H$ is the Hubble rate at the end of inflation, then we can still easily satisfy the dS conjecture. Here, we have added an $\mathcal{O}(1)$ number in front of the TCC bound to quantify our definition of ``reasonable'' mentioned above. Any such refinement to the TCC would not affect its ability to act as a motivation for the dS conjecture. In fact, as shall be explained later on, when trying to derive the TCC from other known facets of string theory such as the SDC or the Weak Gravity Conjecture (WGC) \cite{ArkaniHamed:2006dz, Palti:2017elp}, one is typically led to such refinements of the TCC by $\mathcal{O}(1)$ numbers.

If one does not need the precise version of the TCC, as formulated in \cite{Bedroya:2019snp}, but rather any refinement of it does the job as far as its swampland ambitions are concerned, how was this exact version motivated? As mentioned earlier, the upper limit on the duration of a (quasi)-dS phase of accelerated expansion was derived using the restriction that a TP quantum mode never crosses the Hubble horizon. Since this restriction of not allowing a TP mode to cross the Hubble radius seems to be extremely arbitrary on first sight, the authors of \cite{Bedroya:2019snp} made the assertion that the TCC requires that ``sub-Planckian quantum fluctuations should remain quantum''.  This seems to be a rather different definition of the TCC when compared with its goal of limiting the lifetime of metastable dS phases; however, these two definitions can be shown to be exactly equivalent once \textit{crossing the Hubble horizon} is equated with \textit{classicalization} \cite{Bedroya:2019snp}. This was the main reasoning used to claim that TP modes should always remain quantum and never affect our observations as macroscopic classical perturbations.

On the other hand, in the next section we shall show why the TCC can be refined by $\mathcal{O}(1)$ numbers and yet act as a motivation for the dS conjecture. This naturally brings into question the assertion that a TP mode should never the cross the Hubble horizon. In fact, what we plan to show in this work is that \textit{classicalization does not necessarily require a TP mode to cross the Hubble horizon at all}. If one has an interacting theory, decoherence for a TP mode can occur inside the Hubble radius and therefore, turn it classical. We demonstrate this through one explicit, and yet well-known, example in Sec-3,
and then mention several other possible mechanisms. If one cannot use crossing the Hubble horizon as a necessary criteria for turning classical, there is no sense in demanding that a TP mode should never cross it. Thus, in this paper, we would like to show that one can easily have some criterion by which one puts a bound on the lifetime of an accelerating phase of expansion and yet not require that TP modes remain inside the Hubble horizon. The former requirement is backed up by a lot of evidence which have gathered up in favour of the absence of long-lived metastable dS space \cite{Obied:2018sgi,Ooguri:2018wrx, Dvali:2018fqu, Dvali:2018jhn}, at least for time-independent backgrounds \cite{Dasgupta:2018rtp, Dasgupta:2019gcd, Dasgupta:2019vjn}, in any consistent UV-theory of gravity while the latter has no theoretical footing. Thus, our primary job is to de-link these two definitions of the TCC to show that putting an upper bound on the duration of inflation does not equal to banishing TP modes from decohering and turning classical.  

Another clear example where the TCC does not imply the lack of decoherence  of TP modes is one where perturbations are already classical. The primary scenario for this is warm inflation  \cite{Berera:1995ie,Berera:1996nv}, where the primordial spectrum is dominated by thermal fluctuations. Then, it seems that the only way the TCC might come into play is by limiting the lifetime of metastable dS spaces \cite{Berera:2019zdd} (see also \cite{Das:2019hto,Kamali:2019xnt,Goswami:2019ehb,Brandenberger:2020oav}), which in turn may leave some imprint on the (subdominant) quantum contribution to the power spectrum. Considering this, we compute this contribution looking at a special case with excited states for fluctuations. This would not be the generic warm inflation dynamics, but our main new input is to show an example where the quantum part of the spectrum can be significantly enhanced, compared to the standard BD value, in sync with what happens for cold inflation. One of the motivations for choosing such a Bogolyubov rotation of a BD state as our starting point lies in the fact that, given inflation lasts for a limited extent of time, such states parametrize our ignorance about pre-inflationary dynamics. Although long periods of inflation essentially wash out the effects of such states, as the BD vacuum is a (quantum) attractor \cite{Kaloper:2018zgi}, this is precisely where the TCC (albeit, some refined version of it) would come into play to suggest that there should be some reasonable upper bound on the duration of inflation. Furthermore, a low scale model of inflation, as preferred by the TCC, actually requires many more $e$-foldings to wash away the effects of a modified initial state than what it takes for standard slow-roll models \cite{Kaloper:2018zgi, Kaloper:2002cs}. Given these considerations, we examine the effects of such non-Bunch Davies (NBD) states for warm inflation.

\section{Swampland conjectures \& refinements of the TCC}
Our main goal in this section is to show whether the validity of the dS conjecture necessarily leads us uniquely to the TCC. As we shall find, in fact, the dS conjecture would remain valid as long as we have some refined version of the TCC (as defined in \eqref{RefTCC}). In order to do this, we shall proceed as following: Although the TCC was invoked as a mechanism to motivate the dS conjecture from more general quantum gravity arguments, in this work let us first try to take the opposite approach. Since we are interested in testing the validity of the TCC, it makes sense to try and see how, and if, one can come to this conclusion starting with other well-tested aspects of string theory. In this regard, it was shown in \cite{Brahma:2019vpl} that starting from the SDC, and assuming the species bound \cite{Dvali:2007hz, Dvali:2007wp, Veneziano:2001ah}, one can arrive at the TCC with some assumptions regarding the slow-roll dynamics. However, there was another possibility which was mentioned only in passing, namely that from the same starting point, and with fewer assumptions, one can arrive at a refinement of the TCC, weakening it in the sense of \eqref{RefTCC}. Let us revisit those arguments in order to highlight this fact. Furthermore, it would be clear from our calculations that for any such refined version of the TCC, one would still have bounds on the lifetime of metastable dS states and consequently, the (refined) dS conjecture. 

The SDC tells us that, on parametrically large distances on field space, $|\Delta\varphi|/\Mpl\gg 1$, there appears a tower of light states descending from the UV
\begin{eqnarray}\label{Nstar}
	N_*\(\varphi\) \sim  e^{a \, |\Delta\varphi|/\Mpl}\,,
\end{eqnarray}
with an  $\mathcal{O}(1)$ constant $a > 0$. On the other hand, the species bound states that the effective UV cut-off for a gravitational theory is lowered from $\Mpl$ due the presence of a large number of weakly-interacting particles, as follows:
\begin{eqnarray}\label{UV}
	\Lambda_G \lesssim \frac{\Mpl}{\sqrt{N_*}}\,.
\end{eqnarray}
The number of light states, appearing in the SDC tower, below this cut-off $\Lambda_G$ is then given by
\begin{eqnarray}\label{N}
	N_*\(\varphi\) \sim  \frac{\Lambda_G}{\Mpl}\, e^{a \, |\Delta\varphi|/\Mpl}\,.
\end{eqnarray}

Using \eqref{N} and \eqref{UV}, one can see that the effective cut-off gets lowered exponentially due to large field-excursions, namely
\begin{eqnarray}\label{UV1}
	\Lambda_G \lesssim \Mpl\, e^{-a \, |\Delta\varphi|/(3\Mpl)} \,. 
\end{eqnarray}
The energy density of the universe, characterized by the Hubble parameter $\H$, must remain below this scale for the EFT of inflation to be under perturbative control
\begin{eqnarray}\label{H_bound}
	\H < \Lambda_G  \;\;\Rightarrow \H < \Mpl\, e^{-a\, |\Delta\varphi|/(3\Mpl)}\,.
\end{eqnarray}
Before going on to motivate the TCC from this equation, let us make a few points first. Firstly, neither this reasoning, nor this result, is new. It was this same argument which has also been used to motivate the dS conjecture in \cite{Hebecker:2018vxz}. Eqn~\eqref{H_bound} is, in fact, a generalization of the old Dine-Seiberg argument \cite{Dine:1985he}. More importantly, albeit we have derived this equation starting from the SDC and the species bound, other authors have arrived at the same equation using other aspects of string theory. For instance, it has been shown that one gets such an upper bound starting from the WGC (and its variations) \cite{Cai:2019dzj} as well as from Bousso entropy bounds for quasi-dS spaces \cite{Seo:2019wsh,Sun:2019obt}. All of this is to suggest that one arrives at such a bound quite generally in string theory. 

After arriving at \eqref{H_bound}, and having demonstrated its robustness, the next step is the crucial argument used to derive the TCC.  The number of $e$-folds of inflation is given by
\begin{eqnarray}\label{e-folds1}
	N =  \int_{t_i}^{t_f} H\d t = \int_{\varphi_i}^{\varphi_f}\frac{H}{\dot{\varphi}} \d\varphi\,.
\end{eqnarray}
Using the expression for the rate of change of the Hubble parameter in terms of the scalar field
\begin{eqnarray}\label{Hdot}
	\dot{H}= - \(\frac{\dot{\varphi}^2}{2\Mpl^2}\)\,,
\end{eqnarray}
we can express \eqref{e-folds1} as
\begin{eqnarray}\label{e-folds}
	N \simeq \[\frac{1}{\sqrt{2 \epsilon_H}}\]_{\varphi_i}^{\varphi_f}  \frac{|\Delta\varphi|}{\sqrt{2} \Mpl} = \[\frac{1}{\sqrt{\epsilon_H}}\]_{\varphi_i}^{\varphi_f} \frac{|\Delta\varphi|}{2 \Mpl}
\end{eqnarray}
where $\epsilon_H := -\dot{H}/H^2$ and the square brackets indicate the averaged value of a quantity over the entire field range. 

Eliminating $|\Delta\varphi|$ between Eqns \eqref{H_bound} and \eqref{e-folds} gives us the equation
\begin{eqnarray}\label{TCC_refined}
	N < \(\frac{3}{2 a}\) \[\frac{1}{\sqrt{\epsilon_H}}\]_{\varphi_i}^{\varphi_f} \ln\(\frac{\Mpl}{\H}\)\,.
\end{eqnarray}
This looks like a refinement of the original TCC. Firstly, the $(3/2)$ factor comes from dimensional factors, and in $d$-dimensions, should be replaced by $\(d-1\)/\(d-2\)$. The second major factor arises from the appearance of $\[\frac{1}{\sqrt{\epsilon_H}}\]_{\varphi_i}^{\varphi_f}$ which is certainly bigger than $1$ for any accelerating background. The only way one can recover the original form of the TCC from this expression is if $a$ is big enough to counteract these two factors mentioned above, namely if (in $d$-dimensions)
\begin{eqnarray}\label{a}
	a > \(\frac{d-1}{d-2}\)\, \[\frac{1}{\sqrt{\epsilon_H}}\]_{\varphi_i}^{\varphi_f}\,.
\end{eqnarray}
In order to get an estimate for $a$, we can do several things. For a single Kaluza-Klein tower of modes, we find that $a \sim  \sqrt{\(d-2\)/\(d-1\)}$ and is thus not typically large enough to recover the original TCC. However, to get a more rigorous argument, we can consider, say the (scalar) WGC \cite{Gonzalo:2019gjp} as was done in \cite{Cai:2019dzj}. This case shall be discussed later on in some more detail. Alternatively, one can use entropy arguments to estimate for $a$ \cite{Seo:2019wsh,Sun:2019obt}, or use arguments regarding black holes on dS spacetime \cite{Kehagias:2019iem}, for both of which the exponent turns out to be of the same order of magnitude. However, at first, instead of guessing for $a$, we shall show how it must be necessarily smaller than the lower bound required in \eqref{a}, if one has to recover the refined dS conjecture. Our derivation shall serve the dual purpose of demonstrating that one can have an upper bound on the amount of inflation (or, alternatively, on the lifetime of a metastable dS phase) and yet violate the original TCC constraint. 

Let us return to our trusted Eqn \eqref{H_bound}. Using the Friedmann equation, for a positive potential $V$, one immediately finds that due to the positivity of the kinetic term
\begin{eqnarray}\label{Vbound}
	V_f < \Mpl\, A\, e^{-\frac{2 a |\Delta\varphi|}{3\Mpl}}\,,
\end{eqnarray}
which has already been pointed out in \cite{Bedroya:2019snp}. Here $A$ is a constant depending to the dimension of spacetime and shall be inconsequential for our discussion. As has been shown in \cite{Bedroya:2019snp, Brahma:2019vpl}, one can derive the refined dS conjecture from this expression which includes logarithmic corrections, consequently allowing for short-lived meta-stable dS solutions, when considering all parts of field space. However, in the strict limit of $|\Delta\varphi|/\Mpl \rightarrow \infty$, we are inevitably led to the conclusion that there cannot be any meta-stable dS vacua. In fact, one also gets the original lower bound on the slope of a (positive) potential $V$, given as
\begin{eqnarray}\label{dS_bound}
	\[\frac{|V'|}{V}\]_\infty > \frac{2 a}{3 \Mpl} \sim \mathcal{O}(1)\,,
\end{eqnarray}
where the left hand-side denotes the averaged value of $V$ over asymptotically large field regimes. Plugging in \eqref{a}, we find that for the original version of the TCC to be valid, we get 
\begin{eqnarray}
	\[\sqrt{\epsilon_V}\]_\infty >   \[\frac{1}{\sqrt{2 \epsilon_H}}\]_\infty\,,
\end{eqnarray}
where we assume the usual definition for $\epsilon_V$. In other words, we rule out all quasi-dS spaces if we are to derive the exact original version of the TCC! 

On the other hand, requiring the existence of the dS conjecture does not lead to such dire consequences if one is happy to relax the TCC a bit. For instance, if we now choose $a$ to be such that the $\mathcal{O}(1)$ number on the right hand side of \eqref{dS_bound} is the same as in \cite{Bedroya:2019snp}, then  we find
\begin{eqnarray}\label{TCC_refined_1}
	N <  3\, \ln\(\Mpl/\H\)\,,
\end{eqnarray}
since $a=\sqrt{3/2}$ in this case, and we have $\[\sqrt{\epsilon_V}\]_\infty  > \sqrt{2/3}$, as in \cite{Bedroya:2019snp}. Of course, as has been shown in \cite{Brahma:2019vpl}, this requires using the slow roll relation $\epsilon_H \sim \epsilon_V$. This explicitly shows that a bound on the slope of the potential, as given by the dS conjecture, enforces a bound on the duration of inflation -- just that, it is a milder one than the original TCC. 

From another perspective, without using such slow-roll relations, one can explicitly find such a refinement of the TCC coming from the scalar WGC, as has been shown in \cite{Cai:2019dzj}. In this case, one gets an estimate for the exponent of the SDC $a$ from either the sWGC \cite{Palti:2017elp}:
\begin{eqnarray}
	2 \(\partial_\varphi m\)^2 > m^2\,,
\end{eqnarray}
which implies $a\sim \sqrt{2}/3$ or from the strong version of the sWGC \cite{Gonzalo:2019gjp}:
\begin{eqnarray}
	2 \(V'''\)^2 - V''V'''' > \(V''\)^2\,,
\end{eqnarray}
which leads to $a \sim 1/3$.  Both these cases lead to refinement to the TCC due to $\mathcal{O}(1)$ numbers \cite{Cai:2019dzj}, as was the case for using the dS bound above. 

Finally, there has also been a recent study regarding chaos and complementarity for dS spacetimes \cite{Aalsma:2020aib}. In its application to inflationary backgrounds, the authors find that to have backreaction under control, a statement somewhat equivalent to the TCC can be made but only with a refinement such that
\begin{eqnarray}
	N < \ln(S_\text{dS}) =  2 \ln\(\Mpl/\H\)\,,
\end{eqnarray}
from considerations of the scrambling time for dS spacetime. Thus we find that assuming the TCC leads one to the dS conjecture with a $\mathcal{O}(1)$ lower bound; however, when going in the other direction, multiple stringy arguments suggest that one gets a refined version of the TCC, in the sense of \eqref{RefTCC}. From the perspective of putting an upper bound on the duration of inflation, or the lifetime of a metastable dS state, this is not a big difference at all\footnote{Naturally, it goes without saying that if the TCC is refined by an ${O}(1)$ number, so would the upper limit on the lifetime of metastable dS be, following the arguments of \cite{Bedroya:2019snp}.}. However, if one were to take the requirement of not letting a quantum mode which was $\mathcal{O}(\ell_{\rm Pl})$ not cross the Hubble horizon very seriously, then our arguments show that such a condition is very difficult to achieve from quantum gravity arguments. In fact, this gives us some hints that the Hubble horizon is probably not that special, a sentiment which we shall establish more rigorously in the rest of the paper.

\section{Decoherence of TP modes}
Since there does not seem to be any compelling argument for restricting TP modes from crossing the Hubble radius, it is now pertinent to question the premise that any consistent EFT should not allow TP modes from turning classical. In order to critically examine this claim, let us take the following approach -- we want to show that a TP mode can turn classical even without crossing the Hubble horizon. If such a claim can be established, then there should be no fundamental argument against refining the TCC with some $\mathcal{O}(1)$ number, as has been done for all the other swampland conjectures in the past.

Therefore, we will explore (some of the) possible mechanisms through which a sub-horizon mode can decohere, even when they could have originated from TP scales. Behind the assertion of the TCC that: \textit{Sub-Planckian quantum fluctuations should remain quantum} \cite{Bedroya:2019tba}, lies the assumption that a particular mode will decohere only once it crosses the Hubble horizon. This criteria is more or less accepted, since the phase space region corresponding to a certain mode (that crossed the Hubble horizon during inflation) gets \textit{squeezed}, rendering some resemblance to a classical distribution. However, that does not necessarily check the boxes for other criteria, and in fact, one can even get away with not defining an environment or any interaction with it, which is why it is sometimes referred to as `decoherence without decoherence' \cite{Polarski:1995jg}. For these reasons, at first sight it might seem enough to prevent the classicalization of TP modes by keeping them hidden within the Hubble radius, as stated in the TCC.

However, we would like to argue that in spite of keeping such modes smaller than the Hubble horizon during inflationary expansion (and consequently thereafter), there are a plethora of mechanisms through which the said modes can classicalize. In fact, even without inflation TP modes can be present today at observable scales as a consequence of the expanding Universe. To get an intuitive understanding of the situation, we will present a few concrete examples of how much a TP mode is stretched, and how in every case classical physics is dominant, showing from this standpoint that TP modes come as a part and parcel of an expanding Universe.

\subsection{Expansion of TP modes}
In this subsection we will explore some simple albeit meaningful examples to get a grasp of the length scales we are dealing with when tracking the expansion of TP modes.  We will look at examples at three different energy scales. In each case, at the start of the expansion from the initial energy scale, we will examine a mode that has wavelength just larger than $\ell_{\rm Pl}$, in spirit of what we are calling a TP mode, based on our discussion in the Introduction. For the considerations here, it does not really matter if the wavelength was somewhat smaller or bigger than $\ell_{\rm Pl}$, but to stay clear of the issues discussed in the Introduction of whether a field decomposition makes sense for length scales less than $\ell_{\rm Pl}$, we will treat the wavelength just bigger than $\ell_{\rm Pl}$ and continue to call that the TP mode.  What we will do is for each of three energy scales considered below, we will examine a mode that has wavelength just bigger than $\ell_{\rm Pl}$ and follow its expansion history to determine what its wavelength is today. In all three cases we will see that these initial TP modes have redshifted to present day lengths that are easily within the realm of classical observable physics, even macroscopic physics. To do this, we consider a physical wavelength of a mode and see how it changes from its initial stage $I$,
which will be one of the energy scales we consider, to its final stage $0$ which is the present day.

For the first example, we consider a hypothetical case where there was no inflation at all. The Universe starts at energy scale $T_I \sim M_{\rm Pl}$, and it simply undergoes standard big bang cosmology evolution up to today. Clearly such an expansion history is not sufficient to solve the cosmological
puzzles, but for the sake of argument, let us just see how large can a mode that starts as TP at $I$ expands to today.  This would be given by 
\begin{equation}
	\lambda_0 = \lambda_I \left( \frac{a_0}{a_I} \right) = \lambda_I \frac{T_I}{T_0} \frac{g_* (T_I)^{1/3}}{g_* (T_0)^{1/3}}\;,
\end{equation}
where $g_*$ is the number of degrees of freedom. Hereafter, we neglect the ratio between the number of degrees of freedom as it only introduces a small correction. Thus, the current wavelength of these modes is given by $\lambda_0 = \lambda_I T_I/T_0$, where $T_0 \sim 10^{-31} M_{\rm Pl}$.  For TP modes 
$\lambda_I \sim \ell_{\rm Pl}$, it leads to $\lambda_0 \lesssim 10^{31} \ell_{\rm Pl} \sim 10^{-4}\ {\rm m}$. This means that in this scenario a TP mode would have been stretched to a length roughly the size of a (large) bacteria. Perhaps more importantly, these scales are typical of the thermal radiation emitted by objects at room temperature. The fact that a TP mode can redshift to lengths typical of (classical) electromagnetic radiation highlights 
our point. Even if one excluded any consideration of inflation whatsoever, a mode that was TP at the earliest stage of the Universe will expand, just by standard big bang evolution, to a length scale today in the realm occupied by classical physics.  We will show in the next subsection dynamical mechanisms that can render such modes to be classical, but even without these specifics it seems very plausible just from these length scale considerations that modes expanding to these scales today would somehow have become classical. In a similar fashion to the TCC one could censor a portion of the spectrum as a way to deal with this. However, the futility of such approach is clear for the case at hand.

For the next example, we consider the case of a standard inflation, which \textit{lasts} for $60$ $e$-folds and is followed by a reheating phase with a temperature $T_r \sim M_{\rm GUT}$. Then, 
\begin{equation}\label{eq:str}
	\lambda_0 = \lambda_I\ e^{N_e+N_r}\  \frac{T_r}{T_0} \frac{g_* (T_r)^{1/3}}{g_* (T_0)^{1/3}} \;,
\end{equation}
where the subindex $r$ denotes values at the reheating phase and $I$ stands for the beginning of inflation. Similar to the previous case, we neglect the ratio of the numbers of degrees of freedom, while also ignoring the amount of expansion during reheating. Thus, $\lambda_0 \sim \lambda_I e^{60} T_r/T_0 \sim 10^{57} \lambda_I$, such that for TP modes $\lambda_0 \sim 10^{22}\ {\rm m} \sim 5 \times 10^{10}\ {\rm AU}$, i.e., roughly the size of the largest galaxies! Of course this comes as no surprise, since, crudely speaking, this is the TP problem/window, in the sense that large scales can be traced back to trans-Planckian fluctuations at the beginning of inflation.  As an aside, notice that if we once again ignored inflation and just looked at how large a TP mode at the start of the GUT scale redshifts to today under just big bang expansion, it also leads to a fairly large length scale today, $\lambda_0 \sim 10^{-6}\ {\rm m}$. This is a scale larger than the size of proteins, ribosome and the size of some bacteria (E-coli). Once again, classical processes are ubiquitous at this scale, including some portion of the visible spectrum. 

Finally, we examine the inflationary scenario constrained by TCC in \cite{Bedroya:2019tba}. In that case, the energy scale $V^{1/4} \sim T_r \lesssim 10^{-10} M_{\rm Pl}$, with an amount of expansion during inflation of $N_e \gtrsim 44.4$ $e$-folds (larger amounts of inflation correspond to smaller energy scales). Then, the present day length of modes which were TP at the start of this inflation, would be $\lambda_0 \sim 10^{40}\ \ell_{\rm Pl} \sim 10^5\ {\rm m}$. This is just one order of magnitude less than the typical radius of a planet. Also, antennas can generate waves at these wavelengths for military purposes. Furthermore, notice that this is a cautious prediction, since we have considered the largest scale allowed by TCC (smaller energies lead to more inflationary expansion) and we have neglected the expansion during reheating. This low energy inflation was constructed by \cite{Bedroya:2019tba} to prevent classicalization of TP modes, but as this simple length scale argument shows, TP modes at the start of this inflation would today be at planetary scales, and that alone makes it hard to imagine how they would not have become classical by now.  Nevertheless, if one wished to entertain that possibility, the next subsection outlines various mechanisms that could decohere such modes and make them classical.

\subsection{A mechanism for sub-horizon decoherence for TP modes}
So far, we have just looked at length scales and made the point that TP modes would have stretched to macroscopic scales, even when the TCC is accounted for. One could still argue that this does not necessarily imply a classicalization of those modes. However, there are many subhorizon processes that could turn such modes classical. Some examples are as follows. Mechanisms involving considerable particle production at a given mode may be invoked. The most well-known example in cosmology is (p)reheating \cite{kofman1997towards, Allahverdi:2010xz}. Such a dynamics is pictured to occur at the end of cold inflation as a means to turn the supercooled universe after inflation into one in a thermal state at high temperature in a radiation dominated regime. During (p)reheating the scalar field driving inflation (inflaton) decays into other particle species which then thermalize via collisions. Arguably, a thermal distribution is the example par excellence of classicality. Also, it has been established that particle production by itself can be a decoherence factor \cite{calzetta_decoherence_1990}. There, it was shown that the physical vacuum decoheres only if the expansion leads to particle creation, with a direct relationship between the suppression of interference and the number of created particles. Furthermore, and more relevant to our case, particle creation due to parametric resonance during preheating can also be an effective mechanism for decoherence, as discussed in \cite{Son:1996zs, Shtanov:1994ce}. This sort of (p)reheating process need not follow after an inflationary phase. Similar dynamics might also ensue during some phase transition in the early universe. For our proposes here, the point is it demonstrates an explicit mechanism that classicalizes subhorizon modes of a field.

Taking inspiration from the well-known treatment of reheating as given in \cite{kofman1997towards}, we will show that it is possible to have efficient particle production for modes that were once TP. We will work with a light scalar field $\chi$ and show that modes of it that were once TP can become classical at subhorizon scales during a (p)reheating process. For this, we also need a second (spectator) scalar field $\phi$, that starts oscillating around the minimum of its potential.  As a demonstrative example, we will consider the Electroweak scale. The potential has the form
\begin{equation}
	V(\phi) = \frac{\lambda}{4}(\phi^2 - \sigma^2)^2\,,
\end{equation}
where $\sigma$ is the expectation value of the field. For asymptotically large times, the oscillations take the form
\begin{equation}\label{eq:osc}
	\phi(t) = \Phi(t) \sin(mt), \hspace{1.5cm} \Phi(t) = \frac{\Mpl}{\sqrt{3\pi} mt}\,,
\end{equation}
where $m$ is the mass of the $\phi$ scalar field. Furthermore, we 
assume $\phi$ to be coupled to the light scalar field $\chi$ through the 
interaction Lagrangian ${\cal L}_I \sim -g^2 \phi^2 \chi^2/2$. The $\chi$ 
field is expanded in modes as
\begin{equation}
	\hat{\chi}(\textbf{x},t) = \frac{1}{(2\pi)^{3/2}} \int d^3 k \left(\chi_k^* (t)\, \hat{a}_k \,e^{i\textbf{k}\cdot \textbf{x}} + \chi_k (t)\, \hat{a}_k^{\dagger}\, e^{-i \textbf{k} \cdot \textbf{x}}\right) \;,
\end{equation}
where $k$ denotes the physical momentum. Then, we will interpret the growth of these modes as particles being created at that momentum. This will happen as long as the amplitude of the oscillations of $\phi$ decay faster than shown in \eqref{eq:osc}. Moreover, if linearity holds, then each mode is decoupled and satisfies the equation 
of motion
\begin{equation}
	\ddot{\chi}_k + (k^2 + m_{\chi}^2 + g^2 \Phi^2 \sin^2 (mt)) \chi_k = 0\,.
\end{equation}
Changing the variable to $z= mt$ renders the so-called Mathieau equation:
\begin{equation}
	\chi_k^{\prime\prime} + (A_k - 2 q \cos(2z))\chi_k = 0\,,
\end{equation}
where the constants are given by
\begin{equation}
	A_k = \frac{k^2 + m^2_{\chi}}{m^2} + 2q, \hspace{1.5cm} q = \frac{g^2 \Phi^2}{4m^2}\,.
\end{equation}
The Mathieau equation shows unstable regions that lead to the exponential 
growth of the occupation number of the modes, which increase as $e^{qz}$. 
There are two alternatives in this regard. First, for $q \ll 1$, particle 
production occurs in a narrow band about $k = m$. At the opposite 
end, $q \gg 1$ leads to particle production for a wide range of wavelengths. 
This is known as parametric resonance. 

Other conditions need to be satisfied in order to have particle production; in particular, the adiabaticity condition must be violated, which happens when $\dot{\omega_k} > \omega_k^2$, with
\begin{equation}
	\omega_k^2 = k^2 + m_{\chi}^2 + g^2 \Phi(t)^2 \sin^2 (mt)\,.
\end{equation}
Hence, particle production takes place for modes satisfying
\begin{equation}
	k^2 \leq \frac{2}{3\sqrt{3}} g m \Phi - m_{\chi}^2\,.
\end{equation}
This expression is easily generalized to include the Hubble expansion, which renders
\begin{equation}
	p^2 \equiv \frac{k^2}{a^2} \leq \frac{2}{3\sqrt{3}} g m \Phi - m_{\chi}^2\,,
\end{equation}
where $p$ and $k$ denote the physical and comoving momenta, respectively. One gets access to a broader range of momenta when the background field is roughly
\begin{equation}\label{eq:fmax}
	\phi(t) = \phi_*  \approx \frac{1}{2} \sqrt{\frac{m\Phi}{g}} \approx \frac{1}{3} \Phi q^{-1/4}\,,
\end{equation}
which corresponds to a momentum of
\begin{equation}\label{eq:max}
	p_* \sim \sqrt{g m \Phi} = \sqrt{2} m q^{1/4}\,.
\end{equation}
Then, one can estimate that particle production in this regime takes place 
on a timescale of order $\Delta t_* \sim p_*^{-1}$, roughly the period of 
one oscillation of $\chi$, in agreement with the uncertainty principle. 
The process will be efficient if
\begin{equation}\label{eq:eff}
	qm \gtrsim \Gamma\,, \hspace{1.5cm} q^2 m \gtrsim H\,,
\end{equation}
where $\Gamma$ is the decay rate of the field. Using now
the electroweak energy scale, $H_{\rm EW} \simeq 4.46 \times 10^{-33} \Mpl$, and we will assume that $m \simeq 5 \times 10^{-17} \Mpl$, i.e., 
the Higgs mass. This is a sensible assumption since one expects the 
mass of the particles to be of the order of the temperature at 
that stage, $T_{\rm EW} \sim 10^{-16} \Mpl$. Then, the inequalities 
above can be easily satisfied for any $q > 1$ (and even smaller), which 
is precisely the regime favoured for parametric resonance. The specific 
value of $q$ is not particularly important for our purposes, since for 
a wide range $q^{1/4} \sim {\cal O}(1-10)$, rendering $p_* \sim 10^{-16} \Mpl$. Finally, notice that $p_*^{-1} \ll H_{\rm EW}^{-1}$, which signals how 
the particle production rate is faster than cosmological expansion as a consequence of \eqref{eq:eff}. Thus, decoherence will occur during a Hubble time.

Next, we need to check if TP modes of the $\chi$ field can stretch to the scale we just found.  We will consider the more restrictive case, which is that imposed by TCC in the model of \cite{Bedroya:2019tba}. As mentioned in the previous section and found in \cite{Bedroya:2019tba}, $N_e \simeq 44$ e-folds and $T_r \sim 10^{-10}M_{\rm Pl}$, which implies that by the time of the EW phase transition the TP modes would stretch to $p_{\rm TP} \gtrsim 10^{-33} M_{\rm Pl}$. Clearly, $p_*$ lies in that range, so a TP mode becomes classical due to particle production.

A few comments are in order. First, we have chosen the EW scale and the value of the Higgs mass just for the sake of concreteness. No explicit association is being made here with the Standard Model. However EW is one scale where due to the experimental discovery of the Higgs particle, it is highly suggestive that a phase transition did occur in the early universe. This makes the case for classicalization of modes during the EW transition less hypothetical. Nevertheless, for our example neither $\phi$ nor $\chi$ are assumed to be coupled to any of the SM particles. In this sense, $\chi$ would be mimicking a dark matter field, resembling scenarios explored elsewhere \cite{Falkowski:2012fb,Cline:2012hg,Chowdhury:2014tpa,Shajiee:2018jdq}.  Of course similar arguments could be applied to the actual EW phase transition, where particle production can follow from processes like baryogenesis or even from the gravitational wave environment resulting from bubble nucleation.  

Although the example above was for (p)reheating, there are other dynamical processes in such a phase transition that could also render a mode to become classical. For example, dissipative dynamics similar to that found in warm inflation. In \cite{bartrum2015}, warm inflation type dissipative dynamics was applied to phase transitions. It was shown that such a process could delay spontaneous symmetry breaking during the standard cosmological history and induce late periods of warm inflation, which would further enhance the decoherence effect on field modes, some of which may have been TP modes in the past.

So far we have focused on a specific mechanism for decoherence inside the horizon, but clearly it is not the only one. In a general sense, one needs a system interacting with an environment in order to achieve decoherence, regardless of any length scales considerations. In \cite{Calzetta1995}, it was shown how non-linearities in a quantum theory can lead to classical distributions. There, the quantum fluctuations of the field as well as of the ones it interacts with act as the environment. The same principle may be applied to other scenarios. For instance, classicality of a field configuration can arise following a phase transition \cite{Lombardo2002,Rivers2002}, where the decoherence mechanism again involves the interaction of long-wavelength modes of the order parameter with an environment consisting of higher energy modes of the field itself as well as others. As a consequence, topological defects like domain walls and vortices are formed. Another example of decoherence inside the Hubble radius was given in \cite{Kiefer2007}, where the environment was considered to be a thermal bath of photons. They found that the decoherence rate depends strongly on the coupling to the environment. Hence, decoherence is a consequence of dissipative processes. We will get a sense of this in a different setup in the next section. Finally, another interesting example discussed in the literature is that of decoherence induced by a thermal graviton ensemble 	\cite{Bao2019, Bachlechner:2012dg}, which can also have consequences throughout the entire cosmological history due to potential effects for vacuum decay rates. 

As an aside, note that this analysis also has relevance for recent work looking for signatures of quantum nature which would allow to discriminate between warm and cold inflation in observations. The search for such signatures has been done by looking at violations of Bell's inequalities \cite{maldacena2016model} or specific fingerprints in non-Gaussianities \cite{Green:2020whw}. However, one should notice that it is plausible that none of these scenarios leave purely quantum or classical signatures. Case in point, as argued above, the reheating phase that follows cold inflation can classicalize modes inside the horizon and probably others that re-enter it during the process. Then, not only quantum signatures should be expected from cold inflation, and any endeavour to look for such signatures should account for that fact. A similar argument can be applied for warm inflation. Clearly, in that case the thermal spectrum is dominant, but a quantum contribution can also be accommodated. Furthermore, if the duration of inflation is to be restricted by TCC, then it is natural to expect some enhancement to the quantum bit of the spectrum originated by the excited initial state\footnote{Even in the case of cold inflation, such excited quantum states can also produce similar effects in the expression for the bispectrum (in the folded limit) \cite{Flauger:2013hra, Agarwal:2012mq}, as the ones expected from a classical distribution of perturbations \cite{Green:2020whw}, thus making such a distinction far more murky than typically presumed.}. In the next section we will show that this is in fact the case, while also providing another example of the interplay between quantum and classical perturbations.

\section{Quantum contribution to the primordial power spectrum in warm inflation}
In this section we examine a different role played by TP modes in cosmology. It is understood that warm inflation can accommodate both a quantum as well as a thermal contribution to the primordial power spectrum. Observables are mostly dependent on the dominant thermal term, but in some regimes of low dissipation, remnants of the quantum contribution could manifest themselves. To understand to what degree TP modes affect warm inflation, the regimes where quantum modes play a role need to be identified. 

Since we have shown that there is no pressing need to banish TP modes from crossing the horizon, one can ask what are the remaining takeaways from the TCC? As shown earlier, it seems that even if the TCC gets refined, there would still be some reasonable constraint on the amount of inflation allowed. In fact, this also comes out as a common finding of the swampland conjectures in general that any phase of accelerated expansion must be short-lived. Nevertheless, numerically our refined TCC implies it could still allow for enough $e$-folds of inflation, even at high energy scales, to solve the cosmological puzzles, which is the foremost accomplishment of inflation. However, given this restriction on the lifetime of dS states, it is pertinent to point out that one of the most significant consequence of this would lie in modifying the initial state of perturbations from the BD vacuum\footnote{In the context of cold inflation, another consequence of the swampland is the preference of hilltop-type models \cite{Brahma:2020cpy}.}. In this section, we will outline the computation of a special case where the quantum contribution to the spectrum has excited states for quantum fluctuations. As mentioned in the Introduction, such states arise generically from different considerations of pre-inflationary dynamics such as due to multi-field dynamics \cite{Shiu:2011qw}, cosmological phase transitions \cite{Vilenkin:1982wt}, tunnelling from a false vacuum state \cite{Sugimura:2013cra}, anisptropic expansion \cite{Dey:2011mj}, some non-attractor phase \cite{Lello:2013mfa} or more generally, due to some UV physics \cite{Ashoorioon:2004wd, Schalm:2004qk}. The phenomenological implications for considering such states is well known in the context of cold inflation \cite{Hui:2001ce, Holman:2007na, Agarwal:2012mq, Brahma:2013rua, Flauger:2013hra, Aravind:2014axa, Ashoorioon:2010xg, Ashoorioon:2013eia, Shankaranarayanan:2002ax}. In fact, in the wake of the original TP problem, there were several suggestions pointing to the fact that signatures of TP physics would manifest themselves in the initial state of inflation in their deviation from the BD vacuum \cite{Danielsson:2002kx, Kempf:2001fa, Easther:2002xe, Niemeyer:2000eh}. However, in the context of warm inflation, such states have largely remained unexplored. Firstly, we note that the appearance of such states would be rather natural given some version of the TCC restricting the duration of inflation to not be very large. It is then rather pertinent to consider such states also from the perspective of model-building in warm inflation (since it has eased phenomenological considerations for cold inflation, given the TCC \cite{Brahma:2019unn}). More importantly, \`a priori, one does not know if the quantum and thermal parts of the spectrum would mix with each other in a nontrivial way. In this section, we show that the quantum contribution can indeed be enhanced significantly due to a NBD state while keeping the thermal part unaltered.

For this, we will take a stochastic approach as introduced by Starobinsky for the standard inflationary scenario \cite{Starobinsky:1986fx, Starobinsky:1994bd}, but applied here in the context of warm inflation, building upon the work by Ramos and da Silva \cite{Ramos:2013nsa}. In this program, one separates the perturbation modes into a macroscopic (classical) and a microscopic (quantum) part. As discussed in previous sections, the standard way to split both realms is to compare the scale of interest to the horizon. Then, one can use a suitable window function (see below) which filters the quantum or the classical modes and which includes some parametrization of our ignorance about where exactly the quantum (or the classical) domain begins. In the end, one expects that the results are independent of this parameter. In this sense, in this approach decoherence is imposed (or parametrized) by hand.

First, let us consider the Langevin-type equation of motion of
warm inflation for the full inflaton field, 
\begin{equation}\label{eq:ff}
	\left[\frac{\partial^2}{\partial t^2} + 3H(1+Q) \frac{\partial}{\partial t} - \frac{\nabla^2}{a^2}\right] \Phi + \frac{\partial V(\Phi)}{\partial \Phi} = \xi_T\,,
\end{equation}
where $Q \equiv \Upsilon/(3H)$ is the \textit{dissipative ratio}, $V(\Phi)$ denotes a renormalised effective potential and $\xi_T$ is the noise term which satisfies the fluctuation-dissipation relation 
\begin{equation}
	\left\langle \xi_T (\textbf{x},t) \xi_T (\textbf{x'},t') \right\rangle = 2 \Upsilon T a^{-3} \delta(\textbf{x}-\textbf{x'}) \delta(t-t')\,.
\end{equation}
Next, we decompose the field into a ``short'' and ``long'' wavelength part, such that
\begin{equation}\label{eq:split1}
	\Phi(\textbf{x},t) \longrightarrow \Phi_{>} (\textbf{x},t) + \Phi_{<} (\textbf{x},t)\,.
\end{equation}
Subsequently, we identify the quantum modes with the short wavelength part, and the classical part with the long ones. In this way, one can find the quantum bit through a \textit{window function} such that 
\begin{equation}\label{eq:pq}
	\Phi_{<} (\textbf{x},t) \equiv \phi_q (\textbf{x},t) = \int \frac{d^3 k}{(2\pi)^{3/2}} W(k,t) \left[\phi_{\textbf{k}}(t) e^{-i \textbf{k} \cdot \textbf{x}} \hat{a}_\textbf{k} + \phi^*_{\textbf{k}}(t) e^{i \textbf{k} \cdot \textbf{x}} \hat{a}^{\dagger}_\textbf{k} \right]\,,
\end{equation}
where $W(k,t)$ is the window filter function, which in its most trivial form can be written as
\begin{equation} \label{eq:window}
	W(k,t) = \theta(k-\mu a H)\,,
\end{equation}
with $0 < \mu \ll 1$. This splitting considers every 'long' wavelength mode to be superhorizon, whereas some of the 'short' modes are also superhorizon, but most of them are inside the horizon. The long wavelength modes can be further split into a background contribution, and perturbations on top of it, as
\begin{equation}\label{eq:split2}
	\Phi_{>} (\textbf{x},t) \longrightarrow \phi(t) + \delta \varphi (\textbf{x},t)\,,
\end{equation}
where $\phi(t)$ is the background field and $\delta \varphi (\textbf{x},t)$ represents the classical field perturbation, whose strochastic properties are those that give rise to observable quantities. 

Expanding \eqref{eq:ff} around $\Phi(\textbf{x},t) = \phi(t)$,  with $\delta \phi = \delta \varphi + \phi_q$, yields
\begin{eqnarray}
	&&\frac{\partial^2 \phi}{\partial t^2} + 3H(1+Q)\frac{\partial \phi}{\partial t} + V_{,\phi} \nonumber\\ 
	&& + \left[\frac{\partial^2}{\partial t^2} + 3H(1+Q) \left\{ \frac{\partial}{\partial t}  + H( \theta - \epsilon + \eta) \right\} - \frac{\nabla^2}{a^2} \right] \delta \phi = \xi_T\,,
\end{eqnarray}
where the slow-roll parameters are $\epsilon = -d \ln  H/dN$, $\eta = - d \ln  V_{,\phi}/dN$ and $\theta = d \ln(1+Q)/dN$. We identify the background evolution equation as the first line of the l.h.s., and recover the quantum and classical variables for the field fluctuations, rendering 
\begin{equation}
	\left[ \frac{\partial^2}{\partial t^2} + 3H(1+Q) \left\{ \frac{\partial}{\partial t}  + H( \theta - \epsilon + \eta) \right\} - \frac{\nabla^2}{a^2} \right] \delta \varphi = \xi_q + \xi_T\,,
\end{equation}
where the quantum noise is given by
\begin{equation}\label{eq:nq0}
	\xi_q = - \left[ \frac{\partial^2}{\partial t^2} + 3H(1+Q) \left\{ \frac{\partial}{\partial t}  + H( \theta - \epsilon + \eta) \right\} - \frac{\nabla^2}{a^2} \right] \phi_q\,.
\end{equation}

Let us now go to Fourier space while noticing that, for a de Sitter universe, the conformal time parameter is given by $\tau = -1/(aH)$. For the Fourier modes, the equation for the classical perturbation follows 
\begin{eqnarray}\label{eq:cf}
	\delta \varphi^{\prime \prime}(\mathbf{k}, \tau)-\frac{1}{\tau}(2+3Q) \delta \varphi^{\prime}(\mathbf{k}, \tau)+\left[k^2 + \frac{3}{\tau^2}  (1+Q)(\theta - \epsilon + \eta) \right] \delta \varphi(\mathbf{k}, \tau) \nonumber\\
	=\frac{1}{H^{2} \tau^{2}}\left[\xi_T(\mathbf{k}, \tau)+\xi_{q}(\mathbf{k}, \tau)\right]\,,
\end{eqnarray}
where primes denote derivatives w.r.t $\tau$. Evidently, the quantum noise \eqref{eq:nq0} is now given by
\begin{equation}\label{eq:nq}
	\xi_{q}(\mathbf{k}, \tau)=-H^2 \tau^2 \left[\frac{\partial^2}{\partial \tau^2} - \frac{2+3Q}{\tau}\frac{\partial}{\partial \tau} + \left\{ k^2 +  \frac{3}{\tau^2}  (1+Q)(\theta - \epsilon + \eta) \right\} \right]\hat{\phi}_{q}(\mathbf{k}, \tau)\,,
\end{equation}
where one can easily see from \eqref{eq:pq} that
\begin{equation}\label{eq:pq2}
	\hat{\phi}_{q}(\mathbf{k}, z) = W(k,\tau) \left[\phi_{\textbf{k}} (\tau) \hat{a}_{-\textbf{k}} + \phi_{\textbf{k}}^* (\tau) \hat{a}_{\textbf{k}}^{\dagger}\right]\,.
\end{equation}
As we shall see later on, it will prove useful to express these equations in terms of the dimensionless parameter $z \equiv k/(aH)$, which in the case of the quantum noise renders
\begin{equation}
	\xi_{q}(\mathbf{k}, z)=-H^2 z^2 \left[\frac{\partial^2}{\partial z^2} - \frac{2+3Q}{z}\frac{\partial}{\partial z} + \left\{ 1 +  \frac{3}{z^2}  (1+Q)(\theta - \epsilon + \eta) \right\} \right]\hat{\phi}_{q}(\mathbf{k}, z)\,.
\end{equation}
Plugging \eqref{eq:pq2} into \eqref{eq:nq}, the quantum noise can be written as
\begin{eqnarray}\label{eq:pq3}
	\xi_{q}(\mathbf{k}, \tau) & = & -H^2 \tau^2 \left\{ \left[W^{\prime \prime} (k,\tau) - \frac{2+3Q}{\tau} W^{\prime} (k,\tau) \right] \phi_k (\tau) + 2 W^{\prime} \phi_k^{\prime} (\tau) \right. \nonumber \\
	& & + \left. W(k,\tau) \left[	\phi_k^{\prime \prime} (\tau) - \frac{2+3Q}{\tau} \phi_k^{\prime} (\tau) +  \left\{ k^2 +  \frac{3}{\tau^2}  (1+Q)(\theta - \epsilon + \eta) \right\} \phi_k (\tau)	\right]	\right\} \hat{a}_{-\textbf{k}} \nonumber \\
	& & +\;\; {\rm h.c.}
\end{eqnarray}
From the Fourier decomposition of the full field and its equation of motion \eqref{eq:ff}, the expression inside the brackets in the second line of the equation above vanishes, yielding solutions of the form
\begin{equation}
	\phi_k (\tau) = \tau^{\nu} \left[C_1 (k) J_{\alpha} (k \tau) + C_2 (k) Y_{\alpha} (k\tau) \right]\,,
\end{equation}
where $J_{\alpha}$ and $Y_{\alpha}$ are the Bessel functions of the first and second kind respectively, and
\begin{equation}
	\nu = \frac{3(1+Q)}{2}, \hspace{2 cm} \alpha = \left[\nu^2 - 3(1+Q)(\theta - \epsilon + \eta)\right]^{1/2}\,.
\end{equation}
Consequently, the quantum noise reduces to
\begin{equation}
	\xi_{q}(\mathbf{k}, \tau)  = -H^2 \tau^2  f_{k} (\tau) \hat{a}_{-\textbf{k}} + {\rm h.c.}\,,
\end{equation}
where 
\begin{equation} \label{eq:fk1}
	f_k (\tau) =  \left[W^{\prime \prime} (k,\tau) - \frac{2+3Q}{\tau} W^{\prime} (k,\tau) \right] \phi_k (\tau) + 2 W^{\prime} \phi_k^{\prime} (\tau)\,.
\end{equation}

Finally, the correlation for the quantum noise is conveniently written like
\begin{eqnarray}
	\left\langle \xi_{q}(\mathbf{k}, \tau) \xi_{q}(\mathbf{k}^{\prime}, \tau^{\prime})\right\rangle & = & \left(\tau \tau^{\prime}\right)^{2} H^{4}\left[f_{\mathbf{k}}(\tau) f_{\mathbf{k}^{\prime}}^{*}\left(\tau^{\prime}\right)\left\langle\hat{a}_{-\mathbf{k}} \hat{a}_{\mathbf{k}^{\prime}}^{\dagger}\right\rangle+ f_{\mathbf{k}}^{*}(\tau) f_{\mathbf{k}^{\prime}}\left(\tau^{\prime}\right)\left\langle\hat{a}_{\mathbf{k}^{\prime}}^{\dagger} \hat{a}_{-\mathbf{k}}\right\rangle\right] \nonumber \\
	& = & (2 \pi)^{3} \delta\left(\mathbf{k}+\mathbf{k}^{\prime}\right)\left(\tau \tau^{\prime}\right)^{2} H^{4}[2 n(k)+1] \operatorname{Re}\left[f_{\mathbf{k}}(\tau) f_{\mathbf{k}^{\prime}}^{*}\left(\tau^{\prime}\right)\right]\,,
\end{eqnarray}
where $n(k)$ denotes the statistical distribution of the modes. In particular, one could take the Bose Einstein distribution, such that
\begin{equation}
	n(k) = \left[\exp \left[\frac{k}{aT}\right]-1\right]^{-1}\,.
\end{equation}

\subsection{Stochastic properties}

Observables are related to the emergent stochastic properties of the field perturbation. Specifically, one needs to compute the power spectrum 
\begin{equation}\label{eq:pdv}
	P_{\delta \varphi} = \frac{k^3}{2\pi^2} \int \frac{d^3 k'}{(2\pi)^3} \left\langle \delta \varphi(\textbf{k},\tau) \delta \varphi(\textbf{k}',\tau) \right\rangle\,.
\end{equation}
As advertised before, it is convenient to work with the dimensionless parameter $z = k/(aH)$. In doing so, \eqref{eq:cf} becomes
\begin{eqnarray}\label{eq:cf2}
	&&\delta \varphi^{\prime \prime}(\mathbf{k}, z)-\frac{1}{z}(2+3Q) \delta \varphi^{\prime}(\mathbf{k}, z)+\left[1 + \frac{3}{z^2}  (1+Q)(\theta - \epsilon + \eta) \right] \delta \varphi(\mathbf{k}, z) \nn\\
	&& \hspace{2cm}
	=\frac{1}{H^{2} z^{2}}\left[\xi_T(\mathbf{k}, z)+\xi_{q}(\mathbf{k}, z)\right]\,,
\end{eqnarray}
where the primes now denote derivatives w.r.t $z$. Then, the correlation of the classical perturbation is given by
\begin{equation}\label{eq:corr}
	\begin{aligned}
		\left\langle\delta \varphi(\mathbf{k}, z) \delta \varphi (\mathbf{k}^{\prime}, z)\right\rangle=& \frac{1}{H^{4}} \int_{z}^{\infty} d z_{2} \int_{z}^{\infty} d z_{1} G\left(z, z_{1}\right) G\left(z, z_{2}\right) \frac{\left(z_{1}\right)^{1-2 \nu}}{z_{1}^{2}} \frac{\left(z_{2}\right)^{1-2 \nu}}{z_{2}^{2}}\left\langle \xi_{q}\left(\mathbf{k}, z_{1}\right) \xi_{q}\left(\mathbf{k}^{\prime}, z_{2}\right)\right\rangle \\
		& \hspace{-2cm}+\frac{1}{H^{4}} \int_{z}^{\infty} d z_{2} \int_{z}^{\infty} d z_{1} G\left(z, z_{1}\right) G\left(z, z_{2}\right) \frac{\left(z_{1}\right)^{1-2 \nu}}{z_{1}^{2}} \frac{\left(z_{2}\right)^{1-2 \nu}}{z_{2}^{2}}\left\langle\xi_{T}\left(\mathbf{k}, z_{1}\right) \xi_{T}\left(\mathbf{k}^{\prime}, z_{2}\right)\right\rangle\,,
	\end{aligned}
\end{equation}
where we have expressed the solution of \eqref{eq:cf2} as 
\begin{equation}
	\delta \varphi(\textbf{k},z) = \int_{z}^{\infty} dz' G(z,z') \frac{(z')^{1-2\nu}}{z'^2 H^2} \left[\xi_q (z') + \xi_T (z') \right]\,,
\end{equation}
with its Green's function written as
\begin{equation}\label{eq:green}
	G(z,z') = \frac{\pi}{2} z^{\nu} z'^{\nu} \left[J_{\alpha} (z) Y_{\alpha}(z') - J_{\alpha} (z') Y_{\alpha} (z) \right]\,.
\end{equation}

Observable quantities are related to the comoving curvature perturbation, which in warm inflation receives contributions from the radiation bath and the scalar field, such that
\begin{equation}
	{\cal R} = - \frac{H}{\rho+ p}\left[ \Psi_{\phi} + \Psi_r\right],
\end{equation}
where $\Psi_{\alpha}$ denotes the momentum perturbation (in the spatially-flat gauge) of each species. Numerical \cite{BasteroGil:2011xd} and analytical \cite{Bastero-Gil:2019rsp} studies show that ${\cal R}$ can be well-approximated by
\begin{equation}
	{\cal R} = \frac{H}{\dot{\phi}} \delta \varphi,
\end{equation}
similarly to the cold inflation case. Consequently, the power spectrum is given by 
\begin{equation}
	\Delta^2_{\cal R} = \left(\frac{H}{\dot{\phi}} \right)^2 P_{\delta \varphi},
\end{equation}
where the last term includes the quantum and thermal contributions shown in \eqref{eq:corr}.

From now on we will focus on the quantum bit of \eqref{eq:corr}, specifically focusing on NBD states. We refer the interested reader to \cite{Ramos:2013nsa} for the computation of the thermal part of the spectrum as well as the quantum one using a BD initial state. In our case the thermal contribution to the spectrum remains the same as that computed in that reference. Conversely, \cite{Nezhad:2019mdh} shows the computation of the thermal contribution in the strong dissipative regime by imposing initial conditions at finite past. Since this part of the spectrum is assumed to be purely classical, we consider that the effects of NBD states should only be accounted for in the quantum part of primordial spectrum.

To get back to the point, notice that when working with the variable $z$, the quantum noise term may be written in a similar fashion as in \eqref{eq:pq3}. Case in point,
\begin{equation}
	\xi_q (\textbf{k}, z) = -H^2 z^2 f_k (z) \hat{a}_{-\textbf{k}} + {\rm h.c.}\,,
\end{equation}
where\footnote{Notice that the definitions in \eqref{eq:fk1} and \eqref{eq:fk2} are different.}
\begin{equation}\label{eq:fk2}
	f_k (z) = \left[\frac{\partial^2 W(k,z)}{\partial z^2} - \frac{2+3Q}{z}\frac{\partial W(k,z)}{\partial z} \right] \phi_k (z) + 2 \frac{\partial W(k,z)}{\partial z} \frac{\partial \phi_k (z)}{\partial z}
\end{equation}
and the Fourier component of the field can be expressed as
\begin{equation}\label{eq:pk}
	\phi_k (z) = -i \frac{\sqrt{\pi} z^{\nu} H }{2\sqrt{k^3}} \left[\alpha_k H_{\alpha}^{(1)} (z) - \beta_k H_{\alpha}^{(2)} (z) \right]\,,
\end{equation}
with $H_{\alpha}^{(1)}$ and $H_{\alpha}^{(2)}$ denoting the Hankel functions of the first and second kind, respectively. In the limit $\beta_k \rightarrow 0$, the function corresponds to the BD states. On the contrary, a non-negligible value of $\beta_k$ describes a Bogolyubov rotation of the BD vacuum corresponding to an excited state, i.e., the NBD vacuum.  We shall focus on this case.

The quantum contribution to the power spectrum of the (classical) field perturbation is given by
\begin{equation}\label{eq:pdp}
	P^{\rm qu}_{\delta\varphi} = \frac{k^3}{2\pi^2} |F_k (z)|^2 [2n(k)+1]\,,
\end{equation}
where 
\begin{equation}\label{eq:Fk}
	F_k = \int_z^{\infty} dz' G(z,z') (z')^{1-2\nu} f_k (z')\,.
\end{equation}
Then, our task is reduced to computing $F_k$, since it will easily lead to the quantum contribution to the power spectrum by means of \eqref{eq:pdp}. For this, plug \eqref{eq:green} and \eqref{eq:fk2} into \eqref{eq:Fk}, which renders
\begin{eqnarray}
	F_k & = & \int_{z}^{\infty} dz^{\prime} \left[\frac{\pi}{2} z^{\nu} (z^{\prime})^{\nu} \left(J_{\nu}(z) Y_{\nu}(z^{\prime}) - J_{\nu}(z^{\prime}) Y_{\nu} (z) \right)  \right] (z^{\prime})^{1-2\nu} \nonumber \\
	& & \times \left\{ \left[\frac{\partial^2 W(z^{\prime}-\mu)}{\partial (z^{\prime})^2} - \frac{2+3Q}{z^{\prime}} \frac{\partial W(z^{\prime} - \mu)}{\partial z^{\prime}} \right] \phi_k (z^{\prime}) + 2 \frac{\partial W(z^{\prime} - \mu)}{\partial z^{\prime}} \frac{\partial \phi_k (z^{\prime})}{\partial z^{\prime}} \right\}\,.
\end{eqnarray}
The integral can be performed by using the properties of the Heaviside step function, yielding
\begin{eqnarray} \label{eq:Fk2}
	F_k = - \left. \frac{\partial}{\partial z^{\prime}} \left[G(z,z^{\prime}) (z^{\prime})^{1-2\nu} \phi_k (z^{\prime}) \right] \right|_{z^{\prime} = \mu} + \left. \left[2 \frac{\partial \phi_k (z^{\prime})}{\partial z^{\prime}} - \frac{2+3Q}{z^{\prime}} \phi_k (z^{\prime}) \right] \right|_{z^{\prime} = \mu}\,,
\end{eqnarray}
where the first term on the r.h.s comes from the second derivative of the window function and the last term comes from the first derivative, resulting in a Dirac delta. 

Next, we take the slow-roll approximation at zeroth order, such that $\nu \simeq \alpha$, to then use the following property of the Hankel and Bessel functions (generically denoted as $Z_{\alpha}$),
\begin{equation}
	\frac{d Z_{\alpha} (z)}{dz} = Z_{\alpha-1} (z) - \frac{\alpha}{z}Z_{\alpha} (z)\,.
\end{equation}
Hence, from \eqref{eq:green} and the NBD state \eqref{eq:pk} we get
\begin{equation}
	\frac{\partial G(z,z^{\prime})}{\partial z^{\prime}} = \frac{\pi}{2} z^{\nu} (z^{\prime})^{\nu} \left[J_{\nu} (z) Y_{\nu-1} (z^{\prime}) - J_{\nu-1} (z^{\prime}) Y_{\alpha}(z) \right] + {\cal O}(\nu-\alpha)\,,
\end{equation}
\begin{equation}
	\frac{d \phi_k (z)}{dz} = -i \frac{\sqrt{\pi} z^{\nu} H }{2\sqrt{k^3}} \left[\alpha_k H_{\nu-1}^{(1)} (z) - \beta_k H_{\nu-1}^{(2)} (z) \right] + {\cal O} (\nu-\alpha)\,.
\end{equation}
Then, taking the derivatives in \eqref{eq:Fk2} and replacing the expressions above, we arrive to the simple formula
\begin{eqnarray} \label{eq:Fk21}
	F_{k} (z) & = & -i \frac{\pi^{3/2} z^{\nu} H}{4k^{3/2}} \left\{ \mu \left[\alpha_{k} H_{\nu-1}^{(1)} (\mu)- \beta_{k} H_{\nu-1}^{(2)} (\mu)\right] \left[J_{\nu} (z) Y_{\nu} (\mu) - J_{\nu} (\mu) Y_{\nu} (z) \right] \right. \nonumber \\
	&& \left.- \mu \left[\alpha_{k} H_{\nu}^{(1)} (\mu)- \beta_{k} H_{\nu}^{(2)} (\mu)\right] \left[J_{\nu} (z) Y_{\nu-1} (\mu) - J_{\nu-1} (\mu) Y_{\nu} (z) \right] \right\}\,.
\end{eqnarray}
Expanding around $\mu = 0$ simplifies considerably this expression, yielding
\begin{equation}\label{eq:Fk3}
	F_{k}(z) \simeq -i \frac{\pi^{3/2} z^{\nu} H}{4k^{3/2}} \left(-\frac{2}{\pi}\right) \left[\alpha_k H_{\nu}^{(1)}(z) - \beta_k H_{\nu}^{(2)}(z) + {\cal O}(\mu^{2\nu})\right]\,.
\end{equation}
Furthermore, working at order ${\cal O}(Q)$, such that $\nu = 3/2$ and noticing that for small $z$
\begin{eqnarray}
	H_{3/2}^{(1)} (z) & \simeq & - \sqrt{\frac{2}{\pi z}} e^{iz} \left[1 + \frac{i}{z}\right]\,, \\
	H_{3/2}^{(2)} (z) & \simeq & - \sqrt{\frac{2}{\pi z}} e^{-iz} \left[1 - \frac{i}{z}\right]\,,
\end{eqnarray}
gives the usual non-BD expression for cold inflation, i.e.,
\begin{equation}
	F_k (z) \simeq \frac{-i H}{\sqrt{2k^3}} \left[\alpha_k (z+i) e^{iz} - \beta_k (z - i) e^{-i z} \right]\,.
\end{equation}
This is usually written in terms of the conformal time, such that
\begin{equation}
	F_k (\tau) \simeq \frac{H}{\sqrt{2k^3}} \left[\alpha_k (1+i k \tau) e^{-ik \tau} + \beta_k (1 - i k \tau) e^{i k \tau} \right]\,,
\end{equation}
with $|\alpha_k|^2 - |\beta_k|^2 = 1.$ 

If a vacuum mode $k$ contains $N_k$ particles, and the relative phase between the coefficients is $\theta_k$, then we can write
\begin{equation}
	\alpha_k = \sqrt{1+N_k}\, e^{i \theta_k}\,, \hspace{2cm} \beta_k = \sqrt{N_k}\,.
\end{equation}
In the limit $\tau \rightarrow 0$, we have that
\begin{equation}
	|F_k (\tau)|^2 \simeq \frac{H^2}{2k^3} |\alpha_k + \beta_k|^2 = \frac{H^2}{2k^3} \left[1 + 2N_k + 2 \sqrt{N_k (N_k + 1)}\, \cos \theta_k \right]\,,
\end{equation}
so that the quantum contribution to the power spectrum of the field perturbation is
\begin{equation}\label{eq:pnbd}
	P_{\delta \varphi}^{\rm qu} \simeq \frac{H^2}{4 \pi^2} \left[1 + 2N_k + 2 \sqrt{N_k (N_k + 1)}\, \cos \theta_k \right] \left[1 + 2 n(k)\right]\,.
\end{equation}
As a consequence, the quantum bit of the warm inflation primordial spectrum can be doubly enhanced. On one side, there is a thermal enhancement coming from the statistical distribution of the modes, which may be seen as a thermal influence on the quantum modes. On the other hand, we have found that excited initial states manifest through another enhancement factor similar to that in cold inflation, and it should be considered a purely quantum effect. 

Notice that this analysis at order ${\cal O}(Q)$ already shows the physics of NBD states in a warm inflationary setup. However, one could consider higher order terms in $Q$ in \eqref{eq:Fk3}, which would show some non-trivial relation between dissipation and the quantum part of the spectrum. This would be a dynamical effect that would need to be studied in more detail elsewhere.

\subsection{Another (possible) example of subhorizon decoherence}
As mentioned above, the parameter region defined by $0 < \mu \ll 1$ implies that all the classical modes are larger than the horizon as well as some of the quantum modes (at least for the window function \eqref{eq:window}). However, in the same spirit as the rest of this work, we ask ourselves the consequences of working with $\mu \gg 1$, which would model a scenario where every quantum mode is subhorizon as well as some portion of the classical modes, i.e., decoherence inside the horizon. 

Fortunately, the analysis is rather similar to the previous section, where for the convenience of the reader we re-write \eqref{eq:Fk2},
\begin{eqnarray}
	F_{k} (z) & = & -i \frac{\pi^{3/2} z^{\nu} H}{4k^{3/2}} \left\{ \mu \left[\alpha_{k} H_{\nu-1}^{(1)} (\mu)- \beta_{k} H_{\nu-1}^{(2)} (\mu)\right] \left[J_{\nu} (z) Y_{\nu} (\mu) - J_{\nu} (\mu) Y_{\nu} (z) \right] \right. \nonumber \\
	&& \left.- \mu \left[\alpha_{k} H_{\nu}^{(1)} (\mu)- \beta_{k} H_{\nu}^{(2)} (\mu)\right] \left[J_{\nu} (z) Y_{\nu-1} (\mu) - J_{\nu-1} (\mu) Y_{\nu} (z) \right] \right\}.
\end{eqnarray}
Approximating this for very large values of $\mu$ leads to a similar expression as before, i.e.,
\begin{equation}
	F_k \simeq -i \frac{\pi^{3/2} z^{\nu} H}{4k^{3/2}} \left(-\frac{2}{\pi}\right)\left[\alpha_k H_{\nu}^{(1)}(z) - \beta_k H_{\nu}^{(2)}(z)\right] \left[1 + \mu^{-4} f(Q)\right],
\end{equation}
where 
\begin{equation}
	f(Q) = -\frac{3}{8}Q - \frac{15}{32}Q^2 + {\cal O}(Q^3).
\end{equation}
There are a few conclusions we can get from this. Firstly, at leading order, the power spectrum is independent of the window parameter $\mu$ both for small and larger values of $\mu$. Differences emerge only at higher orders. Case in point, for $\mu \gg 1$ the higher-order terms vary as $\mu^{-4}$, and $f(Q)$ in fact enhances this factor. One could argue that this is just a consequence of a well-known fact, dissipation induces decoherence. A clear argument in favour of this statement is that for cold inflation (or $Q=0$) this term is completely absent. Hence, because warm inflation has dissipation built in, it is another example of decoherence of subhorizon modes. In theory, modes well inside the horizon (larger values of $\mu$) could undergo classicalization so long as dissipation is strong enough. 

On the other hand, consider the parameter region $0 < \mu \ll 1$, where higher-order terms vary as $\mu^{3(1+Q)}$ (eq.\eqref{eq:Fk3}), which could be interpreted as the lack of dissipation-induced decoherence for superhorizon modes. Curiously, this suppression is enhanced by larger values of $Q$. This means that as $Q$ becomes more predominant so do classical processes, which cannot have non-local effects, like the decoherence of superhorizon modes. These are only some qualitative observations. Further consideration would require a dynamical treatment looking at interactions between the classical and quantum processes.

\section{Discussion}
It has been recently argued that macroscopic physics should not arise from modes that were TP at any point in the past. That underlies the TCC condition that no TP mode should cross the Hubble horizon or, in other words, the decoherence time of TP modes should effectively be infinite. Therefore, it is assumed that even if TP modes can be generated, they cannot become classical and thus observable. In this work we have demonstrated this not to be the case. First, from string theory arguments and other swampland conditions, we have shown that the TCC conditions can be relaxed somewhat, to the extent that allows some TP modes to cross the horizon. This loosening of the TCC condition though not significant, is still adequate to make meaningful differences in inflation model building. Moreover, a second point argued here is that even if TP modes remain well inside the horizon, they can still become classical due to interactions throughout the standard cosmological expansion. Thus, the TP classicality condition is simply not consistent with an interacting quantum field theory in an expanding Universe. To emphasize this point, we showed that even without inflation, TP modes can be redshifted up to observable physics scales. Thus, Fourier modes that were TP in the past might be ubiquitous nowadays, and their role is evident regardless of horizon crossing nor from any role of inflation in the early history of the Universe.

Moreover, if TCC is to be understood as the lack of classicalization of TP modes, this rationale should be applied in the same way for other alternatives like string gas cosmology \cite{Brandenberger:1988aj, Brandenberger:2006vv}, ekpyrotic scenario \cite{Khoury:2001wf} and matter bounce cosmology \cite{Finelli:2001sr}. Such scenarios are supposed to be shielded against the TP problem (remaining insensitive to TCC) because cosmological length scales today cannot be traced back to the sub-Planckian realm. However, as we have seen, if TP modes can be stretched enough to make them macroscopic even in the absence of inflation, then the argument of them remaining quantum does not hold. If the TCC is going to stand as a principle of not letting TP modes turn classical, it seems that banishing accelerated expansion is not sufficient to realize such an objective. In fact, if TP modes can become classical even without crossing the Hubble horizon, then there must be a dynamical process that prevents TP modes from being generated at all. In that sense, string gas and bouncing cosmologies recover their appeal, whereas for inflation some other UV-complete theory must step in, like the production of black holes providing a natural mechanism against the generation of TP modes to begin with. 

It should be emphasized that we are not arguing against the restriction on the duration of inflation due to some form of a relaxed TCC condition, as in fact, we have shown how the dS conjecture can naturally lead to such constraints. Actually, a relatively short inflationary period is not in tension with observations \cite{Berera:1997wz,Berera:2000wz}, particularly with respect to the low quadrapole momentum measured by the COBE \cite{Kogut:1996us}, WMAP \cite{Bennett:2012zja} and Planck \cite{Rassat:2014yna} missions. A standard inflationary scenario fails to explain this feature unless one invokes cosmic variance, systematic errors or astrophysical phenomena, and even then, it might not be enough to exhibit a lack of large-angle correlation \cite{Copi:2008hw}. In this sense, a primordial cosmological account may be preferred. This idea has been further explored in \cite{Hirai:2005pg, Hirai:2007ne} and more recently in \cite{Cicoli:2014bja, Das:2015ywa}, where it is pointed out that a minimal duration together with the appropriate dynamics/setup at the beginning of inflation (usually a radiation-dominated era) can yield to a suppressed power spectrum at low $\ell$. The existence of short-lived meta-stable dS spaces and the bound on the duration of inflation from TCC can also have other appealing theoretical features. Firstly, a short period of inflation (close to the lower bound) can follow from anthropic considerations \cite{Freivogel:2005vv}. Moreover, they could be considered as the embodiment of the solution to the graceful exit problem. In the early days of inflation, it was thought that this process should end through quantum tunneling to the minimum of the potential, but it was rather difficult to connect inflation with the radiation dominated era. This was solved by introducing the concept of slow-roll inflation, which ends for values of the field where the potential is too steep. This picture describes inflation taking place on metastable regions (no minimum of the potential considered), in accordance with the restrictions imposed by the swampland conjectures and TCC. Thus, slow-roll and the mechanism to end inflation can be thought of as a natural consequence of short-lived meta-stable dS spaces and the relaxed TCC conditions obtained in this paper. 

Hence, small(er) inflationary periods seem to be more generic, which in turn implies that one should consider NBD initial states, even more so for pre-inflation scenarios with earlier radiation-dominated eras or for multi-stage inflationary scenarios (see \cite{Berera:2019zdd, Mizuno:2019bxy, Li:2019ipk, Cai:2019hge, Dhuria:2019oyf, Torabian:2019zms, Brandenberger:2019eni, Kamali:2019gzr, Marsh:2019bjr} for such setups and similar in the context of TCC). This is common to every inflationary scenario, regardless of the nature of perturbations. Therefore, one should take this into account even in warm inflation, where thermal fluctuations are generally predominant, but for exceptional cases there may be some mixture with quantum fluctuations. Considering this, we have computed the quantum contribution to the power spectrum in warm inflation for the case where there are excited initial states. As expected, at leading order one recovers the same expression as for cold inflation times a generic enhancement term (for warm inflation) coming from a potential statistical distribution of the modes. Further corrections for non-negligible dissipation remain to be explored. Finally, this scenario proved to be another laboratory for the idea of decoherence inside the horizon, which can be possible due to dissipation, one of the key features of warm inflation.

\section*{Acknowledgments}
AB is supported by STFC. SB is supported in part by funds from NSERC, from the Canada Research Chair program, by a McGill Space Institute fellowship and by a generous gift from John Greig. JRC is supported by the Secretary of Higher Education, Science, Technology and Innovation of Ecuador (SENESCYT).

\end{document}